\documentclass[aps,rmp,
			   groupedaddress,superscriptaddress,
			   amsfonts,amssymb,amsmath,
			   citeautoscript,bibnotes,
			   a4paper,showkeys,showpacs]{revtex4}

\setcitestyle{numbers,square,sort&compress} 

\usepackage{microtype} 

\usepackage{txfonts}  
\usepackage{txfontsb} 
\usepackage{bm} 
\usepackage{dsfont} 
\usepackage{textcomp} 
\usepackage{amsmath}
\usepackage{hyperref}
\hypersetup{
    colorlinks,
    linkcolor={blue!75!black!80!yellow},
    citecolor={blue!75!black!80!yellow},
    urlcolor={blue!75!black!80!yellow}
}

\usepackage[centering,hmargin=2.15cm,tmargin=31mm,bmargin=37mm]{geometry}

\usepackage{xcolor}
\usepackage{graphicx}

\newcommand{\appropto}{\mathrel{\vcenter{
  \offinterlineskip\halign{\hfil$##$\cr
    \propto\cr\noalign{\kern.2pt}\sim\cr\noalign{\kern-2.5pt}}}}}
\newcommand{\spar}{{\scriptscriptstyle\parallel}}
\newcommand{\kpar}{k_{\spar}}

\makeatletter
\renewcommand{\fnum@figure}{\figurename~\thefigure\ (color online)}
\makeatother

\usepackage{setspace} 

\begin{document}

\title[Graphene-plasmon polaritons: From fundamental properties to potential applications]{\vskip .5em Graphene-plasmon polaritons: From fundamental properties to potential applications}

\author{Sanshui~Xiao}
\email{saxi@fotonik.dtu.dk}
\affiliation{\scriptsize Department of Photonics Engineering, Technical University of Denmark, DK-2800 Kgs. Lyngby, Denmark}
\affiliation{\scriptsize Center for Nanostructured Graphene, Technical University of Denmark, DK-2800 Kgs. Lyngby, Denmark}

\author{Xiaolong~Zhu}
\affiliation{\scriptsize Department of Micro and Nanotechnology, Technical University of Denmark, DK-2800 Kgs. Lyngby, Denmark}
\author{Bo-Hong~Li}
\affiliation{\scriptsize Department of Photonics Engineering, Technical University of Denmark, DK-2800 Kgs. Lyngby, Denmark}
\affiliation{\scriptsize Center for Nanostructured Graphene, Technical University of Denmark, DK-2800 Kgs. Lyngby, Denmark}
\author{N.~Asger~Mortensen}
\email{asger@mailaps.org}
\affiliation{\scriptsize Department of Photonics Engineering, Technical University of Denmark, DK-2800 Kgs. Lyngby, Denmark}
\affiliation{\scriptsize Center for Nanostructured Graphene, Technical University of Denmark, DK-2800 Kgs. Lyngby, Denmark}

%

\begin{abstract}
With the unique possibilities for controlling light in nanoscale devices, graphene plasmonics has opened new perspectives to the nanophotonics community with potential applications in metamaterials, modulators, photodetectors, and sensors. This paper briefly reviews the recent exciting progress in graphene plasmonics. We begin with a general description for optical properties of graphene, particularly focusing on the dispersion of graphene-plasmon polaritons. The dispersion relation of graphene-plasmon polaritons of spatially extended graphene is expressed in terms of the local response limit with intraband contribution. With this theoretical foundation of graphene-plasmon polaritons, we then discuss recent exciting progress, paying specific attention to the following topics: excitation of graphene plasmon polaritons, electron-phonon interactions in graphene on polar substrates, and tunable graphene plasmonics with applications in modulators and sensors. Finally, we seek to address some of the apparent challenges and promising perspectives of graphene plasmonics.

\end{abstract}


\maketitle

\noindent{\small \textbf{\textsf{CONTENTS}}}\\
\begin{spacing}{1}
{
\begingroup 
\let\bfseries\relax

\tableofcontents
\endgroup
}
\end{spacing}

\section{Introduction}
Plasmonics~\cite{Maierbook,Brongersma:2015a}, as a separate branch of nanophotonics, has recently attracted intensive attention driven by both emerging applications and the maturing of state-of-the-art nanofabrication technology~\cite{Schuller2010,Editorial:2015,Baev:2015}.
With strong interaction between light and free electrons, plasmonics allows breaking the diffraction limit~\cite{Gramotnev:2010,Gramotnev:2014} to concentrate light into deep-subwavelength volumes with huge field enhancements~\cite{Maierbook,Schuller2010,Xiao2011P1,Xiao2010P2,Smith:2015}. The fascinating properties of plasmonics provide the foundation for various applications including integrated optical circuits~\cite{Bozhevolnyi2006,Ansell2015,Xiao2006P1}, single-molecule detection~\cite{Xu1999,Punj2013}, super-resolution imaging~\cite{Kawata2009,Wei2014}, photovoltaic~\cite{Atwater2010,Xiao2012}, color generation~\cite{Kumar2012,Jeppe2014,Zhu:2015}, and biological sensing~\cite{Anker2008,Liu2011}. These achievements were mostly realized in the visible or near-infrared regions where the noble metals are hosting plasmons that can be driven by light fields.

With the rapid rise of graphene, new scientific and technological opportunities are appearing~\cite{Ferrari:2015} and in the context of optoelectronics and photonics, highly doped graphene is considered a promising plasmonic material working in the mid-infrared and terahertz (THz) spectral windows~\cite{Grigorenko2012,Bludov:2013,Abajo2014,Low2014,Vakil2011}.
Compared to plasmon polaritons in noble metals~\cite{Raetherbook}, the graphene-plasmon polariton (GPP) exhibits even stronger mode confinement and relatively longer propagation distance, with an additional unique ability of being electrically or chemically tunable~\cite{Liu2011,Ding2015,Phare2015}. These extraordinary features of graphene plasmons have stimulated an intense line of investigations on both fundamental properties of graphene plasmons~\cite{Koppens2011,Thongrattanasiri2012,Christensen2014} as well as the prospective for applications in metamaterials~\cite{Vakil2011,Lee2012}, modulators~\cite{Sensale2012,Liang2015}, photodetectors~\cite{Ju2011}, and sensors~\cite{Marini2015,Rodrigo2015}. Naturally, plasmon properties depend on the charge-carrier density and high doping levels of graphene can be achieved either by electrostatic top-gating~\cite{Chen2011} or by chemical doping with surface treatment~\cite{Khrapach2012}. Despite the progress in achieving yet higher doping levels, it remains a challenge to push graphene plasmons to the near-infrared or even visible frequency regimes. In this paper, we review recent progress by emphasizing graphene plasmonics from fundamental light-matter interactions to some potential applications. The basic optical properties of non-radiating graphene-plasmon polaritons (GPP) in an extended graphene sheet are firstly reviewed, and then three main subjects of graphene plasmonics are discussed: (1)~Excitation of graphene-plasmon polaritons; (2)~Plasmon-phonon interactions in graphene on polar substrates; (3)~Tunable graphene plasmonics and its applications.
Finally, we provide a brief outlook discussing challenges and perspectives of graphene plasmonics.

\section{Fundamental properties of graphene}
\begin{figure}[!tb]
\includegraphics[scale=0.85]{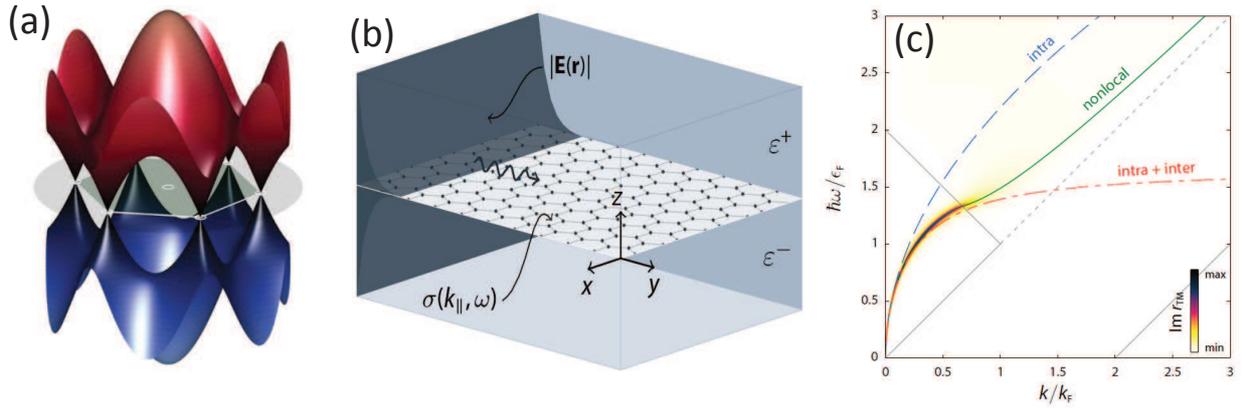}
\caption{(a) The dispersion relation of graphene's electrons within a tight-binding treatment. The first Brillouin zone is indicated in shaded green and delimited by white lines.
(b) A graphene sheet, characterized by its conductivity $\sigma(k_{\parallel},\omega)$, sandwiched between super and substrates with relative dielectric functions $\varepsilon^{\pm}$. (c) Dispersion of GPPs in a free-standing graphene sheet ($\varepsilon^{\pm}=1$) calculated with an analytical local-response intraband description (the dashed blue line), a full local-response conductivity (the dashed-dotted line) accounting for both intra and interband contributions, and finally the full nonlocal conductivity (the green line). Figure is reproduced with permission from ref.~\cite{Thomas2015}.
\label{fig:dispersion}}
\end{figure}
We begin with the electronic bandstructure of graphene calculated in the tight-binding approximation~\cite{Bostwick2007,Castro2009}, which is shown in Fig.~\ref{fig:dispersion} (a). One of the unique properties is that the electron energy follows a linear dispersion $E(k)\approx\hslash v_\textrm{F} k $ near the Dirac point, with Fermi energy $E_\textrm{F}\approx\hslash v_\textrm{F} \sqrt{\pi n}$ where $v_\textrm{F} \approx 10^6$\,m/s is the Fermi velocity and $n$ is the charge-carrier concentration. We note that for pristine graphene in a nearest-neighbor tight-binding description, the unit cell is occupied by two electrons which at low temperatures fully populate the low-energy band [shown in blue color in Fig.~\ref{fig:dispersion}(a)], while the high-energy band (shown in red) remains empty. Thus, undoped graphene ($n=0\Rightarrow E_\textrm{F}=0$) behaves as a semimetal, sometimes also being referred to as a zero-bandgap semiconductor. The optical conductivity of pristine graphene $\sigma=e^2/4\hbar$ is independent of any material parameters~\cite{Nair2008}, giving a constant value of 2.3$\%$ for the optical absorption of a single-layer graphene from the visible to near-infrared spectral windows. This featureless absorption is a manifestation of the massless Dirac fermions and quite remarkably the observed absorption number ties closely with the fine-structure constant $\alpha=e^2/(4\pi\varepsilon_0\hbar c)$, i.e. $\pi\alpha \approx \pi/137\approx 2.3\%$. While it is already fascinating that an atomically thin layer can absorb as much as 2.3\%, it is at least equally attractive that optical properties of graphene can be substantially modified by changing the Fermi energy with the further addition of charge carriers~\cite{Ding2015,Phare2015,Liu2011}.

Plasmons in graphene have many properties in common with longitudinal plasmon oscillations in two-dimensional electron gases supported by semiconductor heterostructures, but it is predicted that graphene also supports a transverse electric mode~\cite{Mikhailov2007}, which does not appear in the more common two-dimensional electron gasses with parabolic energy dispersion. Here, we consider the transverse magnetic mode in an extended graphene sheet (characterized by its conductivity $\sigma(k_{\parallel},\omega)$, where $k_{\parallel}$ is the wave vector of the GPP) sandwiched between two dielectric surroundings, see Fig.~\ref{fig:dispersion} (b). The condition for the transverse magnetic polarized field propagating in the plane can be described as~\cite{Jablan2009,Koppens2011}:
\begin{equation}\label{eq:disperion}
\frac{\varepsilon^+}{\sqrt{\kpar^2-\frac{\varepsilon^+\omega^2}{c^2}}} + \frac{\varepsilon^-}{\sqrt{\kpar^2-\frac{\varepsilon^-\omega^2}{c^2}}} = \frac{\sigma(\kpar,\omega)}{i\varepsilon_0\omega},
\end{equation}
where $\varepsilon_0$ is the vacuum permittivity of free space, while $\varepsilon^\pm$ is the relative dielectric function of the surrounding media. In the nonretarded regime where $\kpar \gg \omega/c$, Eq.~\ref{eq:disperion} can be simplified as
\begin{equation}\label{eq:disperion2}
\omega = \frac{\sigma(\kpar,\omega)}{i\varepsilon_0(\varepsilon^++\varepsilon^-)}\kpar.
\end{equation}
The optical conductivity $\sigma(\kpar,\omega)$ can be treated by the nonlocal density-density response function~\cite{Wunsch2006,Hwang2007}. For the local-response limit, the conductivity  $\sigma=\sigma_\textrm{inter}+\sigma_\textrm{intra}$ has distinct contributions due to both interband and intraband transitions~\cite{Koppens2011,Falkovsky2007}. In the low-temperature limit where the Fermi energy $E_\textrm{F}$ exceeds the thermal energy, i.e. $E_\textrm{F}\gg k_BT$, the expressions are:
\begin{equation}\label{eq:conductivity}
\begin{split}
&\sigma_\textrm{intra}(\omega)=\frac{i e^2 E_\textrm{F}}{\pi \hbar^2 (\omega+i\gamma)}, \\
&\sigma_\textrm{inter}(\omega)=\frac{e^2}{4\hbar}\left[\theta(\hbar \omega-2E_\textrm{F})+\frac{1}{\pi}\ln\left |\frac{2E_\textrm{F}-\hbar\omega}{2E_\textrm{F}+\hbar\omega} \right|     \right]
\end{split}
\end{equation}
where $\gamma$ is the relaxation rate and $\theta(x)$ is the Heaviside function. When considering the intraband transition solely, one obtains the following plasmon dispersion relation:
\begin{equation}\label{eq:disperion3}
\omega = \sqrt{\frac{e^2E_\textrm{F}}{\pi\hbar^2\varepsilon_0(\varepsilon^++\varepsilon^-)}\kpar}\Longleftrightarrow \frac{\hbar\omega}{E_\textrm{F}} = \sqrt{2\alpha\frac{c}{v_\textrm{F}}\frac{2}{  \varepsilon^++\varepsilon^-}\frac{\kpar}{k_\textrm{F}}}.
\end{equation}
This generic square-root dispersion relation is referred to as the local-response limit, since in the derivation we have neglected nonlocal response (spatial dispersion) by treating the conductivity itself in the small-wavevector limit, rather than maintaining the full wavevector dependence~\cite{Raza:2015}. From the dimensionless form (right-hand expression), it is also interesting to note that there is a 'universal' dispersion relation irrespectively of the particular doping level, see the dashed blue line in Fig.~\ref{fig:dispersion}~(c).

In the zero-loss and zero-temperature limits, the dispersion of graphene surface-plasmon polaritons, see Fig.~\ref{fig:dispersion} (c), is obtained by three different models: the local-response intraband description, the full local-response conductivity, and the full nonlocal conductivity. Here, we have for the purpose of illustration considered typical numbers given by $E_\textrm{F}=0.4$~eV and $\hbar\gamma=12$~meV.

\section{Excitation of graphene-plasmon polaritons}
\begin{figure}[!tb]
\includegraphics[scale=0.7]{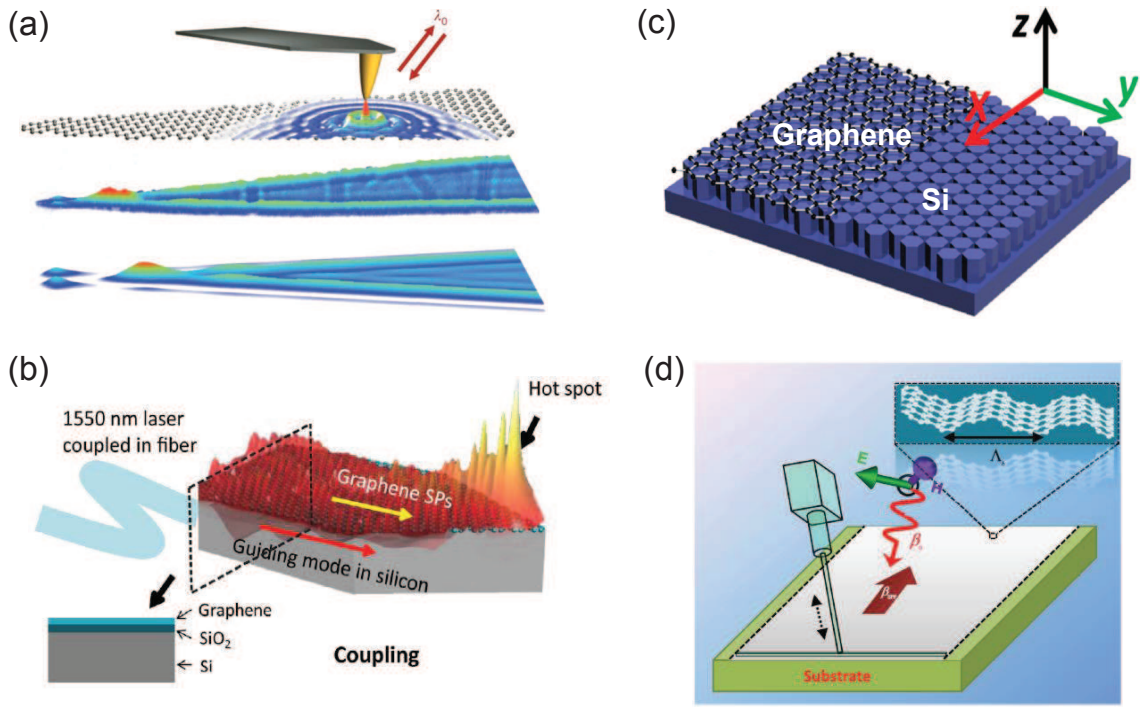}
\caption{(a) Excitation and detection of GPP by use of s-SNOM with a metal-coated tip (in yellow). From top to bottom: Schematic of experimental setup, near-field amplitude image, Colour-scale
image of the calculated local density of optical states. (b) Schematic of experimental configuration to excite GPP with the aid of the guiding mode in Si waveguide. The hot spot appearing at the tip position of the tapered geometry shows the characteristic of GPP.
(c) Illustration of graphene covered nanopatterned Si substrate which provides a wavevector compensation for GPPs by grating coupling. (d) Geometry of 3D graphene grating driven by external acoustic stimulation. The inset shows a sinusoidal grating. Figures are reproduced with permission from ref.~\cite{Chen2012} (a), ref.~\cite{Zhang2014} (b), ref.~\cite{Zhu2013P2} (c), ref.~\cite{Farhat2013} (d).
}
\label{fig:excitation}
\end{figure}
In order to exploit and manipulate GPPs within various nanophotonic and optoelectronic applications, it is crucial to facilitate an efficient coupling of photons to GPPs for the development of future devices. Here, the non-radiating nature of GPPs constitute both a strength and a challenge. Graphene plasmons in nano-/micro-patterned graphene structures such as ribbons~\cite{Ju2011} and disks~\cite{Yan2012}, so-called localized GPPs, can be easily excited. But the strong spatial confinement of GPPs limits their applications when long propagation length is needed for e.g. communication and information processing techniques. Here we pay special attention to propagating GPPs with the challenge to be excited due to the mismatch of the wavevectors of the GPP and of light in vacuum. In this section, we review recently developed technologies/methods for the excitation of propagating GPPs.

The scattering-type scanning near-field optical microscopy (s-SNOM) was almost simultaneously proposed to detect and image propagating GPPs in two Nature papers published in the same issue~\cite{Chen2012,Fei2012}. Chen~\emph{et~al.} launched and detected propagating GPPs in tapered graphene nanostructures using an s-SNOM equipped with a metal coated AFM tip with infrared excitation light, see Fig.~\ref{fig:excitation} (a). In their setup, the metal coated cantilever tip serves as an antenna, which converts the incident infrared light to GPPs with a wavelength of about 40 times shorter than that in free space, see the middle plot in Fig.~\ref{fig:excitation} (a). Excited graphene plasmons are reflected at the ribbon edges, thus producing interference fringes recorded with the same AFM tip. Similarly, Fei~\emph{et~al.}~\cite{Fei2012} worked with graphene on SiO$_2$ and excited GPPs using a slightly different incident wavelength compared to ref.~\cite{Chen2012}, where phonon excitation in SiO$_2$ is avoided~\cite{Fei2011}. The pattern of the calculated local density of optical states, see the bottom of Fig.~\ref{fig:excitation} (a), matches quite well with the image for the GPP produced by s-SNOM. Both papers provide real-space imaging of GPPs with extremely high spatial resolution, around 20~nm limited only by the dimensions of AFM probe. The s-SNOM technology has been intensively used, e.g. exploration of plasmons in graphene moir\'e
superlattices~\cite{Ni2015}, observation of ultraslow hyperbolic polariton propagation~\cite{Yoxall2015}, and observation of hyperbolic phonon-polaritons in boron nitride~\cite{Li2015}. More recently, Alonso-Gonz\'alez~\emph{et~al.}~\cite{Alonso2014} utilized an immovable metal gold rod standing on a graphene sheet as another kind of antenna. Near-field oscillation around the antenna was observed by use of a pseudo-heterodyne interferometer and corroborates the existence of GPPs. Here, a normal Si tip was used, compared to its counterpart with a metal coating layer~\cite{Chen2012,Fei2012}.

Butt-coupling technology widely used in nanophotonics is another way to couple light into GPPs. The coupling efficiency for the butt-coupling is mainly determined by the overlap of the modes. Due to the extreme mode confinement for GPPs, one really needs to
suppress the size of the optical mode in order to have efficient coupling. Nikitin~\emph{et~ al.}~\cite{Nikitin2014} numerically proposed the efficient coupling by the compression of surface polaritons on tapered bulk slabs of both polar and doped semiconductor materials. The infrared photons can be compressed from several micrometers to around 200~nm, allowing to couple light to GPPs with around 25\% conversion efficiency.
Shown in Fig.~\ref{fig:excitation} (b), Zhang~\emph{et~al.}~\cite{Zhang2014} used a guiding mode supported by a silicon waveguide to excite plasmons in graphene on top of silicon, where the graphene is doped by surface carrier transfer method with molybdenum trioxides. The observation of the hot spot in the tip, see Fig.~\ref{fig:excitation} (b), by near-field scanning optical microscopy verifies the existence of GPPs in the near-infrared region.

In the plasmonic community, the grating coupler is widely used to compensate the mismatch between the wavevector of plasmons in the metal and the wavevector of free-space radiation. A similar concept from plasmonic crystals was introduced to excite THz GPPs on a periodic hole array in a graphene sheet by Kitty~\emph{et~al.}~\cite{Kitty2014}. Zhu~\emph{et~al.}~\cite{Zhu2013P2} proposed to use a two-dimensional subwavelength silicon grating beneath graphene to excite GPPs, see Fig.~\ref{fig:excitation} (c). With the aid of dielectric grating, the transmission dips were observed by Fourier transform infrared spectroscopy (FTIR), indicating the existence and the excitation of GPPs in an extended graphene sheet. In the same line, the excitation of GPPs by use of a one-dimensional silicon grating was also investigated theoretically~\cite{Gao2012} and experimentally~\cite{Gao2013}. Graphene is regarded as a perfect two-dimensional material. However, a three-dimensional graphene grating would be possible. By combining a proper substrate and applying acoustic mode excitation, see Fig.~\ref{fig:excitation} (d), Farhat~\emph{et~al.}~\cite{Farhat2013} and Schiefele~\emph{et~al.}~\cite{Schiefele2013} proposed independently to excite GPPs on curved graphene structures. More recent work on curved graphene geometries include refs.~\cite{Christensen:2014b,Wang:2015b,Reserbat-Plantey:2015,Smirnova:2015}.

\section{Plasmon-phonon interactions in graphene on a polar substrate}
\begin{figure}[!tb]
\includegraphics[scale=.8]{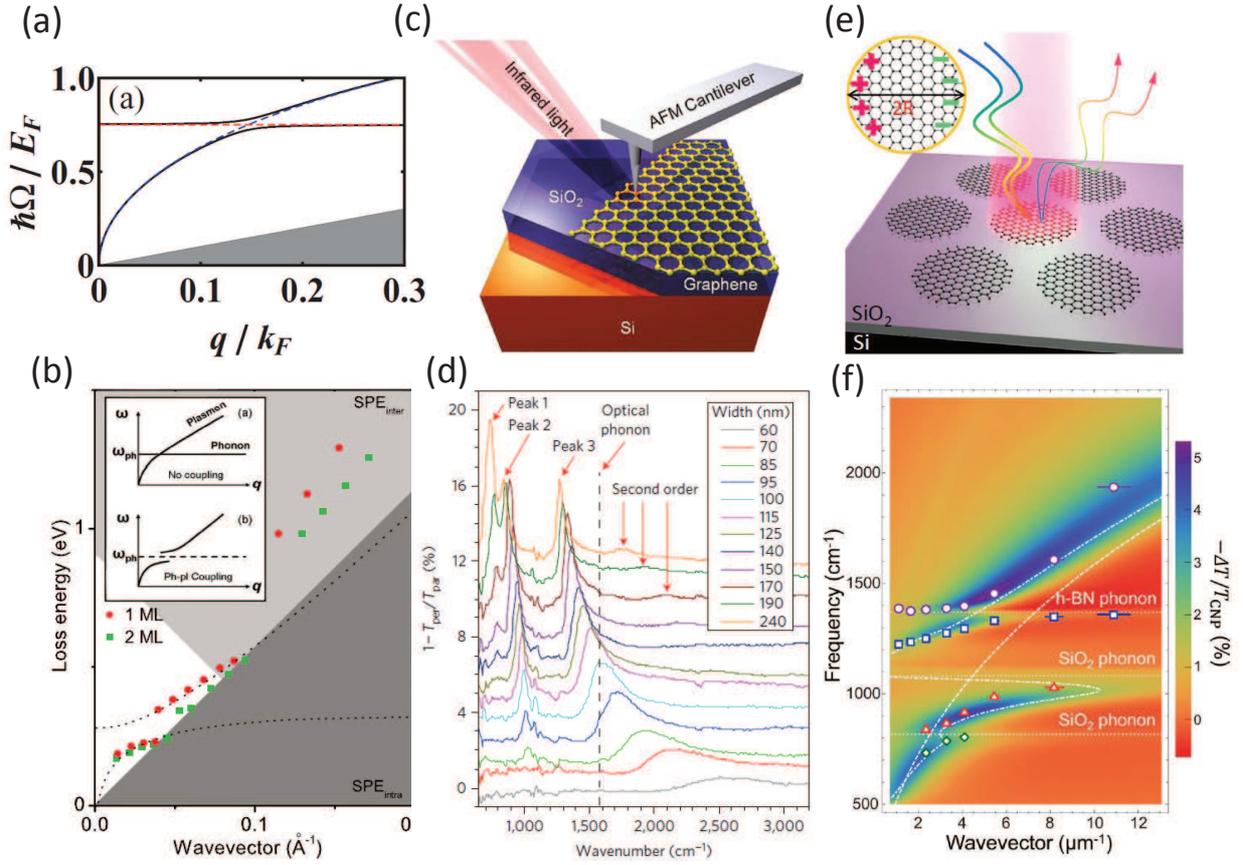}
\caption{(a) Dispersion of hybrid plasmon-phonon modes (the solid lines) and of the uncoupled modes (the dashed lines). Gray areas denote the region of single-particle damping. (b) The dispersion measured by angle-resolved reflection electron-energy-loss spectroscopy in single and multilayer graphene which couple strongly to surface optical phonon of SiC.  (c) Schematics of the infrared nanoscopy (assisted by a nanoscale tip) to study Dirac plasmons at the graphene SiO$_2$ interface. (d) Extinction spectra of graphene ribbons on SiO$_2$ with different ribbon widths. (e) Schematics of
mid-infrared reflection measurement by FTIR microscopy, where the inset shows the dipole oscillation in a graphene dot.  (f) Hybridization of graphene plasmons and phonons in a monolayer h-BN sheet and SiO$_2$. Symbols represent experimental data, and the the dash-dot line indicates the calculated dispersion for graphene/SiO$_2$.
Images reproduced with permission from {ref.~\cite{Jablan2011}}~(a), {ref.~\cite{Liu2010}}~(b), {ref.~\cite{Fei2011}}~(c), {ref.~\cite{Yan2013}}~(d), {ref.~\cite{Zhu2014}}~(e) and {ref.~\cite{Victor2014}}~(f).
\label{fig:phonon}}
\end{figure}
The electron mobility in a suspended single-layer graphene can be as high as 200\,000~cm$^2$V$^{-1}$s$^{-1}$. However, the carrier mobility drops significantly when placing graphene on an insulating dielectric substrate such as SiO$_2$~\cite{Bolotin2008}. The effects of polarizable substrates on carrier dynamics in graphene have been studied~\cite{Fratini2008}, and one of the factors limiting the carrier mobility in graphene supported by a dielectric substrate is the surface optical phonon scattering~\cite{Hess1979}. Due to weak phonon scattering, hexagonal boron nitride (h-BN) is believed to be a promising dielectric material for graphene~\cite{Dean2010}, compared to other substrates such as SiO$_2$ and SiC. When the energy of plasmons becomes comparable with that of phonons, the hybridization of plasmon and phonon modes occurs, leading to the breakdown of the Born--Oppenheimer approximation~\cite{Pisana2007}. The interactions of plasmons with optical phonons in semiconductors have been studied~\cite{Mooradian1996}. In this part, we will review recent studies of plasmon-phonon coupling in graphene, including both the theoretical investigation and experimental demonstrations.

In 2010, Hwang~\emph{et~al.} calculated collective excitation of coupled electron-phonon systems by taking into account the Coulomb coupling between electronic excitations in graphene and the optical phonon modes in the substrate~\cite{Hwang2010}. The dispersion of the coupled plasmon-phonon modes shows that the plasmon-phonon coupling is strong for all electron densities. By using the self-consistent linear response formalism, Jablan~\emph{et~al.}, predicted the existence of coupled plasmon-phonon excitations in graphene~\cite{Jablan2011}, which is shown in Fig.~\ref{fig:phonon} (a). The unique electron-phonon interaction in graphene results in unconventional mixing of plasmon and optical phonons, and the hybridization becomes stronger for larger doping values in graphene.

Strong phonon-plasmon coupled modes in graphene/SiC systems are confirmed using angle-resolved reflection electron-energy-loss spectroscopy~\cite{Liu2010,Koch2010}, where
graphene were prepared on a 6H-SiC crystalline wafer surface by solid-state graphitization~\cite{Forbeaux1998}.
Fig.~\ref{fig:phonon} (b) shows the dispersion behavior of the loss peaks taken on graphene of various thicknesses.
The dispersion looses its continuity both for one-monolayer (red dot in Fig.~\ref{fig:phonon} (b)) and two-monolayer graphene (green squares), where crossovers occur between plasmons and surface optical phonons. The insert in Fig.~\ref{fig:phonon} (b) shows a schematic change of the dispersion curve where two modes couple with each other. The s-SNOM was used to probe the mid-infrared plasmons in graphene~\cite{Chen2012,Fei2012}, and similar technology was also be explored to study plasmon-phonon interactions at the graphene-SiO$_2$ interface~\cite{Fei2011}, see
Fig.~\ref{fig:phonon} (c). The beam of an infrared laser is focused on the metalized tip of an AFM cantilever.
The strong near-field confinement of mid-infrared radiation at the tip apex leads to high spatial resolution and high coupling that enables to investigate the spectroscopic signature of plasmon-phonon interactions. The plasmon-phonon interaction and hybridization at the graphene-SiO$_2$ interface is observed by near-field nanoscopy.

Plasmon-phonon interactions in nanostructured graphene on SiO$_2$ substrate were also studied by FTIR~\cite{Zhu2014,Yan2013}.
In ref.~\cite{Zhu2014}, Zhu~\emph{et~al.} demonstrated an effective approach for patterning graphene sheets into large-area ordered graphene nanostructures by combining nanosphere lithography with O$_2$ reactive ion etching, without high-cost and low-throughput lithography and sophisticated instruments. Relying on nanosphere lithography technology, we have demonstrated stretch-tunable plasmonic structures, broadband enhancement of spontaneous emission, and high light-extraction enhancement~\cite{Ou2014,Zhu2012,Zhu2012P2,Zhu2012P3,Zhu2012P4}.
A schematic of the FTIR mid-infrared reflection measurement scheme is illustrated in Fig.~\ref{fig:phonon} (e). This setup is used to probe the plasmon-phonon interactions in graphene. The hybridization of graphene plasmons and substrate phonons was experimentally demonstrated, showing coupling energies of the order 20~meV. Extinction spectra for graphene nanoribbons
are shown in Fig.~\ref{fig:phonon} (d), illustrating three
main peaks~\cite{Yan2013}.
The nanoribbons with dimensions as small as 60~nm were realized using electron-beam lithography. The multiple features shown in Fig.~\ref{fig:phonon} (d) are ascribed to plasmon–phonon coupled modes. More importantly, they provided a new damping mechanism for GPP with the plasmon lifetime of 20~fs. The surface polar phonons in the SiO$_2$ substrate under graphene nanostructures not only modify the plasmon dispersion, but also provide new channels for damping of GPPw~\cite{Yan2013}.

Experiments have also been extended to investigate the coupling between graphene plasmons and surface phonons in thin polar substrates~\cite{Li2014,Victor2014,Barcelos2015}. The interactions between graphene plasmons and thin layers of PMMA were studied, showing that the PMMA phonon signature is strongly enhanced through graphene plasmon coupling~\cite{Li2014}. Recent FTIR  measurements~\cite{Victor2014,Barcelos2015} verify the hybridization of graphene plasmons and phonons in a
monolayer h-BN sheet. As an example~\cite{Victor2014}, Fig.~\ref{fig:phonon} (f) illustrates the dispersion of graphene plasmons coupled with surface phonons, where experimental data are plotted as symbols. The three horizontal dotted lines indicate the optical phonon energies of h-BN and SiO$_2$, and the dash-dot line shows the coupled dispersion of graphene/SiO$_2$. In addition to the coupling between graphene plasmon and phonons from SiO$_2$, the interaction between graphene plasmons and phonons in h-BN sheet is also clearly observed, showing an avoided crossing (anticrossing) behavior near the h-BN optical phonon at 1370~cm$^{-1}$.

\section{Tunable graphene plasmonics}
Tunable plasmonic materials are essential in various areas of photonics, with applications ranging from perfect absorbers, high-tech frequency modulators and radiators, to highly efficient chemical and bio-chemical sensing~\cite{Ju2011,Brar2013,Jadidi2015,Ding2015,Liu2011,Marini2015,Rodrigo2015,Hedayati2014,Franklin2015,Yang2015,Dyer2013,Fluegel2007}. Graphene supports plasmons that are tunable, providing a novel platform for tunable devices. The large mobility of charge carriers in graphene makes high-speed tunable plasmonics possible by electrical gating of graphene.

\begin{figure}[!tb]
\includegraphics[scale=0.65]{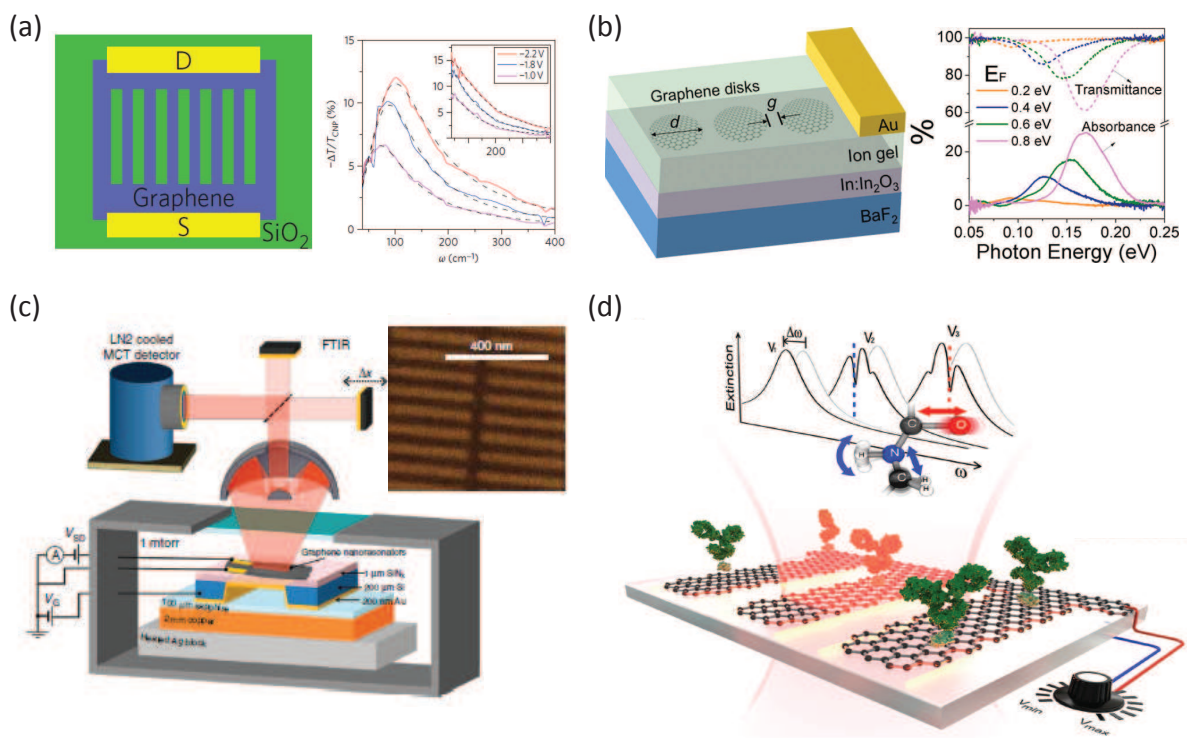}
\caption{(a) Graphene plasmonics for tunable terahertz
metamaterials. Left, illustration of electric back gating of the graphene micro-ribbons. Right, Control of terahertz plasmon excitations through electrical gating. (b) Active tunable absorption enhancement with graphene nanodisk arrays. Left, scheme of the measured devices. Right, FTIR measurements for graphene nanodisk arrays with the tuning of Fermi level by electric top gating. (c) Set-up for electronic modulation of infrared radiation in graphene plasmonic resonators. Inset shows a SEM image of the graphene structure. (d) Plasmonic biosensing with tunable graphene plasmonics implemented in the Mid-infrared frequencies. Images reproduced with permission from {ref.~\cite{Ju2011}}~(a), {ref.~\cite{Fang2013}}~(b), {ref.~\cite{Brar2015}}~(c) and {ref.~\cite{Rodrigo2015}}~(d).}
\label{fig:tunable}
\end{figure}

In 2011, Ju~\emph{et~al.} reported plasmon excitations in graphene microribbons within the terahertz frequency range. As shown in Fig.~\ref{fig:tunable}(a), they demonstrated that graphene plasmon resonances can be tuned over a broad  frequency range by changing the ribbon width and \emph{in situ} back-gating~\cite{Ju2011}. As in the studies of two-dimensional massless Dirac electrons, the graphene plasmon frequency also demonstrates a power-law characteristic behavior. In particular, the graphene plasmon resonances have remarkably large oscillator strengths, resulting in prominent room-temperature optical absorption peaks while the plasmon absorption in a conventional two-dimensional electron gas was observed only at low temperature. Thus, graphene plasmons hold exceptional promise for optoelectronic applications by improving weak optical absorption at visible and infrared wavelengths. In 2013, Brar~\emph{et~al.} demonstrated highly confined tunable mid-infrared plasmonics in graphene nanoresonators~\cite{Brar2013}. They mapped the wavevector-dependent dispersion relations for graphene plasmons at mid-infrared energies from measurements of resonant frequency changes by tuning the charge-carrier density.
In Fig.~\ref{fig:tunable}(b) we an example where Fang~\emph{et~al.} found significantly enhanced graphene absorption (30\%) in arrays of graphene nanodisks by using a In−In$_{2}$O$_{3}$/BaF$_{2}$ substrate and e-beam lithography patterning~\cite{Fang2013}. By controlling the Fermi level of graphene (top-gated ion-gel), a wide range of plasmon tunability in the IR was demonstrated in device geometries.

Dynamic control of thermal radiation through~\emph{in situ} modification of material emissivity could enable the design of novel infrared sources; however, the spectral characteristics of the radiated
power are dictated by the electromagnetic energy density and emissivity, which are usually given properties of the material and the particular temperature.
Except for enhancing the light absorption, dynamic control of the photon radiation could also be enabled by tunable graphene plasmonics. As shown by Brar~\emph{et~al.} (Fig.~\ref{fig:tunable}(c)), the graphene resonators produced antenna-coupled
black-body radiation, which manifested as narrow spectral emission peaks in the mid-infrared~\cite{Brar2015}. Through continuously tuning of the nanoresonator charge-carrier density, the frequency and intensity of these spectral features can be modulated via an electrostatic doping.

Biomaterials have abundant spectroscopic features (phonon vibration modes) in the mid-infrared frequencies~\cite{Li2014}. However, these modes poorly interact with the infrared light. By exploiting the unique electro-optical properties of GPP, it is possible to realize a high-sensitivity graphene-based biosensor for chemically specific label-free detection of proteins.
A mid-infrared GPP based biosensor, see Fig.~\ref{fig:tunable}(d), was reported by Rodrigo~\emph{et~ al.}~\cite{Rodrigo2015}. The tunability of graphene allows to selectively probe the protein at different frequencies. They showed superior sensitivity in the detection of their refractive index and vibrational fingerprints because of the high overlap between the strongly localized GPP mode and the nanometric biomolecules. In fact, from Eq.~\eqref{eq:disperion3} it follows that the resonance wavelength shift $\Delta \lambda$ associated with a refractive-index change $\Delta n$, i.e. the refractometric sensitivity $\Delta\lambda/\Delta n$, is given by
\begin{equation}
\frac{\Delta\lambda}{\Delta n} \simeq  f\times\frac{\lambda}{n},\quad f=\frac{n^2}{n^2 +\varepsilon_s}
\end{equation}
where $\varepsilon^+=n^2$, with $n$ being the refractive index of the sensing medium above the graphene, while $\varepsilon^-=\varepsilon_s$ is the substrate dielectric function. Here, $f$ quantifies the relative light-overlap with the sensing medium in a similar fashion as it has been introduced for dielectric refractive-index sensors based on e.g. photonic crystals~\cite{Mortensen2008}. In the context of the sensing figure-of-merit, ${\rm FOM}= (Q/\lambda)\,
\Delta\lambda/\Delta n$~\cite{Sherry,Jeppesen}, we thus get ${\rm FOM}\simeq (\omega/\gamma)\, n/(n^2 +\varepsilon_s)$. Clearly, the prospects for low GPP damping will benefit sensors through an increased resonance quality factor $Q\simeq\omega/\gamma$.

Moreover, photocurrent in graphene can be harnessed by tunable intrinsic plasmons. Freitag~\emph{et~al.} demonstrated polarization-sensitive and gate-tunable photodetection in graphene nanoribbon arrays~\cite{Freitag2013}. The highly gate tunable graphene plasmon-phonon modes are longer lived and very promising for  plasmonic enhanced photodetectors, exceeding an order of magnitude as compared with excitation of electron–hole pairs alone. The general principle of tunable graphene plasmons can open a new way to engineering photodetectors from terahertz to infrared frequencies.

It is worth mentioning that the combination of graphene with traditional plasmonic structures or metamaterials could also provide the desired dielectric environment with an adjustable feature, resulting in high-speed tunable optical responses. In particular, interband and intraband transitions under optical excitation in graphene have been investigated. In the visible/near infrared range, interband transitions result in a constant absorption of $\sim$2.3\% by single-layer graphene at normal incidence and modulation of infrared absorption by electrically tuning its Fermi level, the absorption can be further improved by applying graphene in designed optical cavities~\cite{Xizhu2013, Ding2015}. In the infrared and terahertz range, intraband transitions dominate, resulting in an optical conductivity well described by the Drude model~\cite{Sensale2012}. Liang~\emph{et~al.} demonstrated that a  monolithically  integrated graphene modulator can efficiently modulate the light intensity with a 100\% modulation depth for certain region of the pumping current at the terahertz range~\cite{Liang2015}.
At infrared frequencies, electrical control of a plasmonic resonance using large-area graphene was demonstrated based on the graphene-covered plasmonic nanowire~\cite{Kim2012} and metasurface systems~\cite{Mousavi2013}, respectively. The combination of plasmonic structures with the possibility to control the Fermi level of graphene, paves the way for advanced optoelectronic devices at optical frequencies.

\section{Discussion and outlook}
Graphene plasmons are commonly explored at mid-infrared and THz frequencies. In order to integrate graphene-plasmon concepts with existing more mature technologies, we need to extend plasmon frequencies into the telecommunication or visible-frequency windows. In graphene nanostructures (with $D$ being the characteristic dimension), the plasmon-resonance frequency associated with localized graphene plasmons scales as $\sqrt{E_\textrm{F}/D}$, which follows immediately from Eq.~\eqref{eq:disperion3}. Thus, higher frequencies can be achieved through both higher doping levels and/or reduced structure dimensions. Graphene plasmons were observed around 3.7~$\mu$m in graphene nanorings with a 20~nm width~\cite{Fang2013}, i.e. with dimensions close to the limits of state-of-the-art electron-beam lithography. A natural way of pushing plasmons to even higher energies is to use bottom-up approaches, e.g., colloidal chemistry methods, or self-assembly of organic molecules~\cite{Cai2010,Li2008,Rasappa2015}. As an example, Cai~\emph{et~al.} presented their bottom-up approach to control the graphene nanoribbons with atomic resolution~\cite{Cai2010}. When the size of graphene structures becomes smaller and smaller, we eventually also need to be concerned with atomic-scale details, such as quantum mechanical effects associated with electronic edge states inevitably hosted by zigzag terminations of the graphene lattice~\cite{Christensen2014,Thongrattanasiri2012,Wang:2015}, which give rise to shift and broadening of plasmon resonances.

It was predicted theoretically that graphene plasmons exhibit relatively long propagation distances. Experimentally, graphene plasmon losses are however still not fully up to this potential. The s-SNOM~\cite{Chen2011} was used to estimate plasmon-propagation distances with the aid of interference fringes in the near vicinity of graphene edges, suggesting a relaxation time in graphene of 30~fs. Recently, relaxation times as large as 500~fs have been observed in high-quality graphene sandwiched between two films of h-BN~\cite{Woessner2015}, which paves a promising way to the development of graphene nanophotonics and nano-optoelectronic devices.
Wafer-scale single-crystalline graphene monolayers are considered as an ideal platform for electronics and other applications.
Single-crystalline graphene monolayer can be achieved by mechanical exfoliation~\cite{Novoselov2004,Zhang2005}, however, still with very small area.  With the recently rapid development of chemical-vapour deposition, inch-sized single-crystalline graphene can be synthesized with a fast growth speed~\cite{Hao2013,Wu2015}. The access to efficient production of wafer-scale monolayers may push graphene closer to real applications.

Finally, we note that here we have only toughed on the linear-response properties of graphene-plasmon polaritons, while graphene may also host rich nonlinear dynamics~\cite{Cheng:2014,Peres:2014,Smirnova:2014,Christensen:2015,Cox:2014,Cox:2015}.

\begin{acknowledgments}
We thank Thomas Christensen and Nicolas Stenger for stimulating discussions. The Center for Nanostructured Graphene (CNG) is sponsored by the Danish National Research Foundation, Project DNRF103.
We also acknowledge the Villum Foundation (341/300-123012) and the Danish Council for Independent Research (FNU 1323-00087).
\end{acknowledgments}

\bibliographystyle{apsrev4-1}
%

\end{document}